\documentclass[aps,amsmath,amssymb,showpacs]{revtex4}

\usepackage{graphicx}
\usepackage{epsfig}

\begin{document}

\title{On ``the authentic damping mechanism'' of the phonon damping model. II
\footnote{I cannot not neither confirm nor disprove that an article \cite{m} 
was rejected by PRC as a reader learns from \cite{nn}.
It is common practice that a manuscript submission to any scientific 
journal remains a confidential information of the author(s) until the 
manuscript 
is accepted for publication. In PRC, it is protected by a secret accession 
code. It seems to me at least not ethical to publish openly on Web 
a confidential information of another person(s) as Mr. N. Dinh Dang did 
in \cite{nn}. An editorial decision may be published only with a formal 
permission from the Editors. I doubt that Mr. N. Dinh Dang has it.
I also interpret this fact as an attempt to influence a reader by
non-scientific arguments in a scientific discussion.}}

\author{V.~Yu.~Ponomarev\cite{byline2}}
\affiliation{Institut f\"ur Kernphysik, Technische Universit\"at Darmstadt,
D--64289  Darmstadt, Germany}
\date{\today }

\begin{abstract}
This article continues a discussion raised in previous publications
(LANL preprint server, nucl-th/0202006 and nucl-th/0202020).
I try to convince my opponents that general arguments are not ``my case"
and may be applied to their model.
\end{abstract}

 
\maketitle

\vspace*{4mm}

To remind in brief a discussion which is already distributed over several
publications: 

A damping mechanism of giant resonances (GR) is well established and
represents now basic knowledge in nuclear structure physics. Calculations
performed by many groups of authors within different microscopic approaches
confirm that a spreading width (due to a coupling of collective modes,
phonons, to complex configurations) is the main part of the total GR width
in medium and heavy nuclei. In light nuclei, a coupling to continuum
(an escape width) also plays an essential role.

A damping mechanism of GRs in a phenomenological phonon damping model (PDM) in
its PDM-1 version is different from that 
(see, an important clarification in \cite{note0}). 
A collective phonon fragments within PDM-1 
as a result of coupling to simple and not to complex  
configurations, i.e. only the so-called Landau damping mechanism is 
accounted for.
A coupling strength is a phenomenological model parameter which is adjusted
to reproduce the GR width known from experiment. Agreement with data
provided by fits within the PDM may be defined as from very good to 
excellent.

In a recent article \cite{n9} which raised the present discussion, it has 
been concluded that these type of fits confirm ``the {\bf authentic} damping 
mechanism'' of the PDM as ``the result of coupling between
collective phonon and non-collective $p$-$h$ configurations''
(i.e. the well established knowledge on the GR properties was put in doubt). 
This conclusion has been criticized in my article 
\cite{m}. It has been argued that this model has the Breit-Wigner (BW)
form for the phonon distribution as an {\it ad hoc} input and thus, even
excellent description of the data available is not surprising.
A fruitfulness of an idea to make conclusions from fits in which model 
parameters are adjusted to described physical observables has been put in 
doubts.

Although my evaluation of the PDM in \cite{m} was made for the point
of view of general physical grounds, Dang {\it et al.} did not agree
with me in the forthcoming publication \cite{nn}. They claim that I consider
some specific case (``his case") which cannot be attached to the PDM and all 
my arguments ``{\it are either wrong or irrelevant}".  
I cannot agree with their conclusion and present below additional arguments
in a sequence following the paragraphs in \cite{nn}:

\vspace*{3mm}

{\bf 2.} For the giant dipole resonance (GDR), 
the energy scale associated with 
variations in a coupling matrix between a phonon and uncorrelated $1p1h$
states is of the order of a few hundred keV.
The width of the GDR strength function is of the order of
a few MeV. So, I do not agree that the condition cited in \cite{nn} from 
\cite{Boh69} is satisfied in the GDR region: why are a few MeV small compared 
to a few hundred keV?

I know only one PDM-1 article \cite{n1} in which it is assumed that a
phonon interacts 40 times stronger with some specific configurations 
than with other ones (see more on this article in {\bf 9.} below). 
In all other PDM-1 papers, we find a single phonon which interacts equally
with all $1p1h$ configurations.
I do not want to discuss here the PDM fits at non-zero temperature.
To keep on reproducing the data in hot nuclear, Dang {\it et al.} have to
assume for unclear reasons that a phonon prefers to interact with $1p1p$
and $1h1h$ configurations about 10 times stronger than with $1p1h$
configurations. Again, as in the case of cold nuclei, an idea to provide
the best fits is preferred to understanding of the physics.
I think it is a blind way for theory.

It is true that PDM equations are presented in a general form in many 
papers by this group with different $V_{q_1 s_1}$. But the point is that
they are never used in actual calculations in this form.  
For this reason, I prefer to discuss what is used in
calculations rather than what is written and not used even by the PDM 
authors themselves.


{\bf 3.} It is very simple to transform Eq.~(1) in \cite{nn} for $m_q^{(2)}$ 
into Eq.~(1) in \cite{m} for $W_2$, although Dang {\it et al.} claim
it is impossible. For that, one needs to switch off an additional PDM 
smearing, i.e., consider the limit $\varepsilon \to 0$. This would bring
immediately to the first line of Eq.~(2D-14) in \cite{Boh69}.
Eq.~(1) in \cite{m} (for a constant coupling strength) or its general
form in \cite{Boh69}: 
\[
\hspace*{60mm}W_2 = \sum_{a,\alpha} (V_{a \alpha})^2 \hspace*{60mm}
\mbox{(2D-14)}
\]
for the second moment $W_2$ is relevant to the PDM as well as to any model
which deals with interactive systems.

Of course, to perform this transformation one should use the PDM
strength function introduced in Ref.~\cite{o1}:
\begin{equation} 
S_q(E) = \frac{1}{\pi} 
\frac{\gamma_q(E)}{\left(E-\omega_{q}-P_q(E)\right)^2+\gamma_q^2(E)}~
\label{e1}
\end{equation} 
where $\gamma_q(E)$ is the PDM damping, $P_q(E)$ is the 
polarization operator (see, e.g., Ref.~\cite{o1} for definitions), and
$\omega_{q}$ is a phonon energy, a model parameter.
The strength function $S_q(E)$ presents fragmentation properties of a PDM
phonon over eigen-states of the PDM Hamiltonian smeared with an additional
parameter $\varepsilon$. Parameter $\varepsilon$ appears in
$\delta(E) = \varepsilon/[\pi \cdot (E^2+ \varepsilon^2)]$ for
$\delta$-functions in $\gamma_q(E)$.

I point this out because the strength function (\ref{e1}) has been
replaced in the forthcoming PDM articles
\cite{n9,nn,n1,n2,n3,n4,n5,n6,n7,n8,n10,n11,n12,n13},
by its approximate form:
\begin{equation} 
S_q'(E) = \frac{1}{\pi} 
\frac{\gamma_q(E)}{\left(E-E_{GDR}\right)^2+\gamma_q^2(E)}
\label{e2}
\end{equation} 
where $E_{GDR}$ should be taken as a solution of
\begin{equation} 
f(E) \equiv E-\omega_{q}-P_q(E)=0~.
\label{e3}
\end{equation} 
Eq.~(\ref{e2}) has been obtained from Eq.~(\ref{e1}) by expanding $P_q(E)$
near a solution of Eq.~(\ref{e3}), $E_{GDR}$, and then extrapolating the 
properties of this approximation far away from $E_{GDR}$.
In the limit $\varepsilon \to 0$, Eq.~(\ref{e3}) has $N+1$ solutions 
corresponding to eigen-energies of the PDM Hamiltonian.

\vspace*{3mm}

{\bf 4.} I never claimed that the BW 
form for the phonon distribution
is assumed within the PDM. But it is indeed an {\it ad hoc} input for PDM
calculations. 
I may refer again to \cite{Boh69} where we read that ``{\it the
Breit-Wigner form for the strength function is an immediate consequence of
the assumption of a constant coupling to the other degrees of freedom of the
system}". The BW under discussion has nothing to do with definition of
the PDM strength function. Indeed, in the limit
$\varepsilon \to 0$, $S_q(E)$ turns into a set of infinitely narrow lines 
while their envelope still remains the BW.

\vspace*{3mm}

{\bf 5.} I do not agree that the calculation with 
random values of $E_{\alpha}$ 
in \cite{m} ``{\it no longer corresponds to the PDM}". 
I have used the PDM Hamiltonian
and details on a spectrum and model parameters are only technical
details of a calculation. The purpose of my calculation is to demonstrate
that ``{\it the crucial feature of the PDM is the use of realistic
single-particle energies}" \cite{nn}
is of marginal importance when a configuration space
is not small; everything is determined by the BW discussed above.

$E_0$ in Ref.~\cite{nn}belongs to the Lorentz line in a hypothetical 
nucleus and not to my 
PDM fits. Eigen energies in my calculation in Ref.~\cite{m}
were obtained from Eq.~(\ref{e3}) in the limit $\varepsilon \to 0$.
 
\vspace*{3mm}

{\bf 6.} I agree that if something ``{\it is by no mean[s] obvious}" 
it has to be
checked. My experience of microscopic calculations tells me that the increase 
of collectivity tends to the increase of a coupling strength. Of course, it
is not necessary that everybody should trust my experience. But then, there
are no other alternatives: the one, who puts it in doubt, should check it 
independently.

\vspace*{3mm}

{\bf 7.} I never claimed that there are some reasons ``{\it why the values
of $f_1$ for $^{40}$Ca and $^{48}$Ca should be the same}" as there are no
reasons to keep this parameter fixed along chains of isotopes. As pointed
out in \cite{m}, this parameter has no physical meaning. 
My issue is that one cannot learn anything from agreement with experiment from
fits in which a free parameter is adjusted to a described observable.
It is important to stress once again:

\vspace*{2mm}

\fbox{
\parbox[50,0]{170mm}{
\textbf{\noindent 
The Phonon Damping Model (PDM)
has three phenomenological parameters, $\omega_q$, $f_1$, and $c_1$
for  position, width, and amplitude of the GDR, respectively, and the
Breit-Wigner as an {\it ad hoc} input for its shape. Numerical values of these 
parameters cannot be determined from independent data.} 
}}

\vspace*{2mm}

On the page 4 of the article under discussion \cite{n9}, we find: 
``{\it For double closed-shell nuclei $^{16}$O and $^{40,48}$Ca, where the
pairing gap is zero, such a kind of enlargement of configuration space is
compensated simply by a renormalization of $f_1$, which reduces its value by
$\sim 25\%$ for  $^{16}$O, and $\sim 35-37\%$ for $^{40,48}$Ca.}" 
I read this statement after discussion of open-shell nuclei as 
a renormalization of $f_1$ in closed-shell nuclei in respect to open-shell
nuclei. Or do Dang {\it et al.} 
mean to say that calculations in double-magic nuclei
have been performed with pairing as one may conclude now from \cite{nn}:
``{\it The results for GDR in $^{16}$O have been obtained already within the
enlarged space}"?
Obviously, only the authors know whether they renormalized their 
$f_1$ in calculations along chains or not. 
Of course, I will take out the statement on the $f_1$ renormalization from 
$^{16}$O to $^{18}$O if it is not true. But
before that, I need some help from the authors 
as to how a reader should interpret the above cited statements.

\vspace*{3mm}

{\bf 8.} It is clearly explained in \cite{m} why comparing the PDM 
predictions in $^{40}$Ca to the data \cite{Har00}, 
the strongest $1^-$ state observed should be excluded from consideration
(because it has a two-phonon nature and two-phonon states are not 
included in the PDM model space). 
Thus, the PDM 0.25\% of the TRK EWS corresponds to 
0.007\% and not to 0.025\% from this experiment. It seems to me that Dang
{\it et al.} try to hide again a huge disagreement by misleading comparison.

The same conclusion, that the PDM is not capable to reproduce 
``{\it the significant experimental difference in the E1 strengths}" 
 $^{40}$Ca/$^{48}$Ca, has been obtained independently by another
group of authors \cite{Har01}. As they write, ``{\it It is 
important to note that
the parameters of the PDM are adjusted to reproduce the gross structure of 
the GDR while investigations of $\gamma$-ray strength function models show 
that the extrapolation of the strength distribution down to energies below 
the particle threshold leads to unrealistic high dipole strengths and 
overestimates the experimental data}".
Thus, it is not only my point of view that a conclusion in \cite{n9}
(on a quantitative description of pygmy dipole resonance within the PDM)
is not justified.

It is not true that the PDM with a structureless phonon has no problems
with double counting. If a phonon internal structure is not accounted for, 
it does not change the physical meaning of the phonon.
The PDM configuration space contains a phonon and 
uncorrelated $1p1h$ configurations. The last ones are also excited from 
the ground state and each of them has its own $B_{1p1h}(E1)$ value. 
If $1p1h$ spectrum is rather complete (it is always true in the PDM 
calculations) these uncorrelated $1p1h$ state alone exhaust 
about 100\% of the TRK EWSR.
But the PDM physics is determined only by the phonon strength function and not
by its sum with $N$ strength functions of uncorrelated $1p1h$ configurations.
This is equivalent to $B_{1p1h}(E1) = 0$ within the PDM.

\vspace*{3mm}

{\bf 9.} The previous article on pygmy resonances 
by Dang {\it et al.} \cite{n1} 
was not a subject of \cite{m}. But if  Dang {\it et al.}  
raise discussion on it in \cite{nn}, I have some comments on it too:

The capability of the phonon damping model (PDM) to describe giant dipole 
resonance (GDR) damping in neutron-rich nuclei has been tested in \cite{n1}. 
To mimic essential differences of double-magic and exotic nuclei,
a coupling between a phonon and some $1p1h$ configurations near the
Fermi surface has been strongly enhanced in the last ones.
As a result, a phonon interacts with these selected $1p1h$ configurations
with a strength ``{\it equal to 41~MeV for oxygen isotopes, 13.856~MeV for
calcium isotopes, and 6.928~MeV for tin isotopes}" \cite{note1}.
Let us try to understand how it is possible to stand such 
an enormous coupling strength which is far away from nuclear structure scales 
and report an agreement with experiment from this type of calculations.
For that, I have repeated the PDM calculations for $^{16}$O and $^{18}$O
at zero temperature keeping all the details of Ref.~\cite{n1}
and employing realistic $1p1h$ spectrum from Hartree-Fock calculation with
SGII Skyrme forces \cite{note2}.
 
\begin{figure}
\epsfig{figure=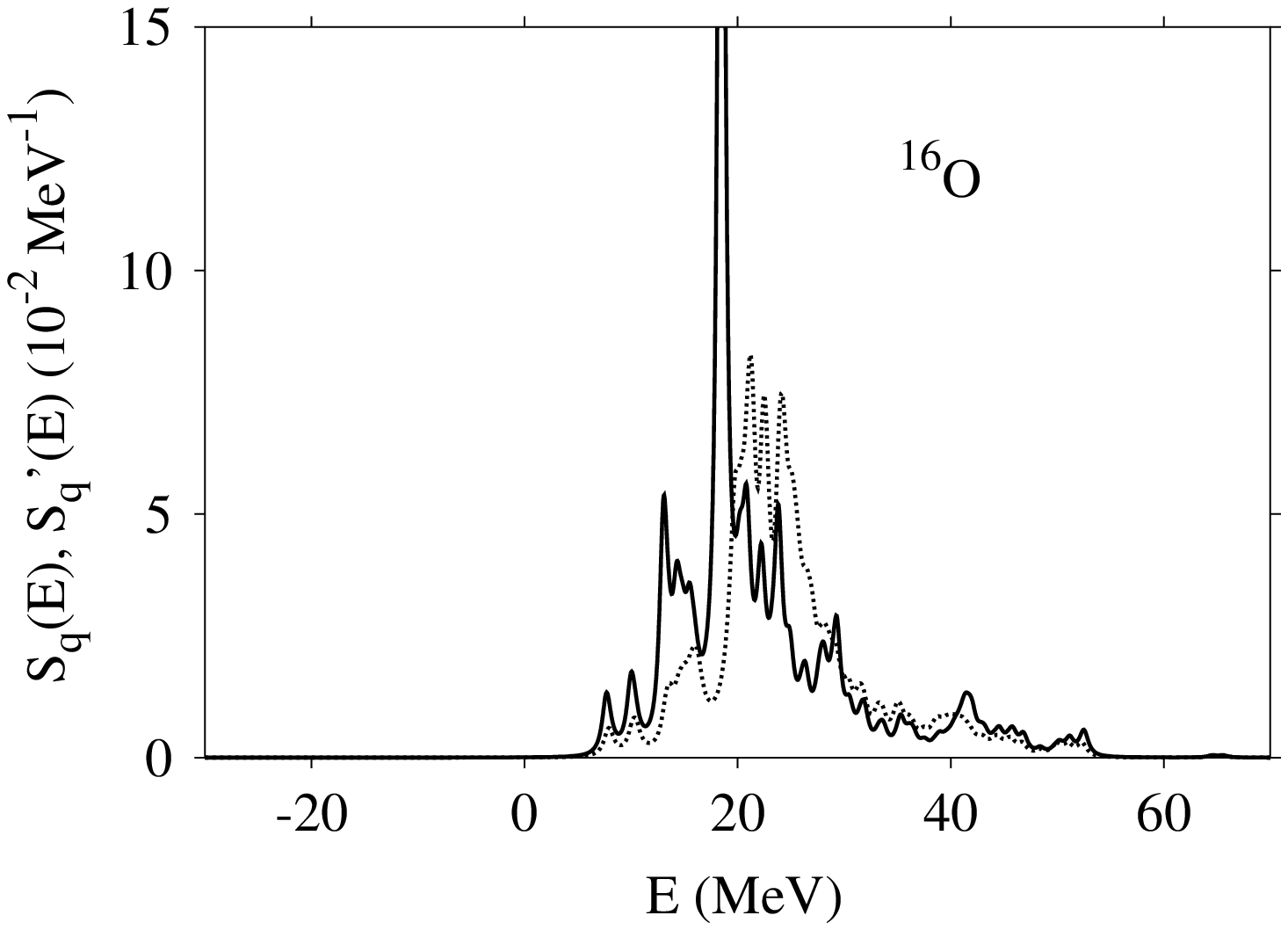,width=85mm,angle=0}
\epsfig{figure=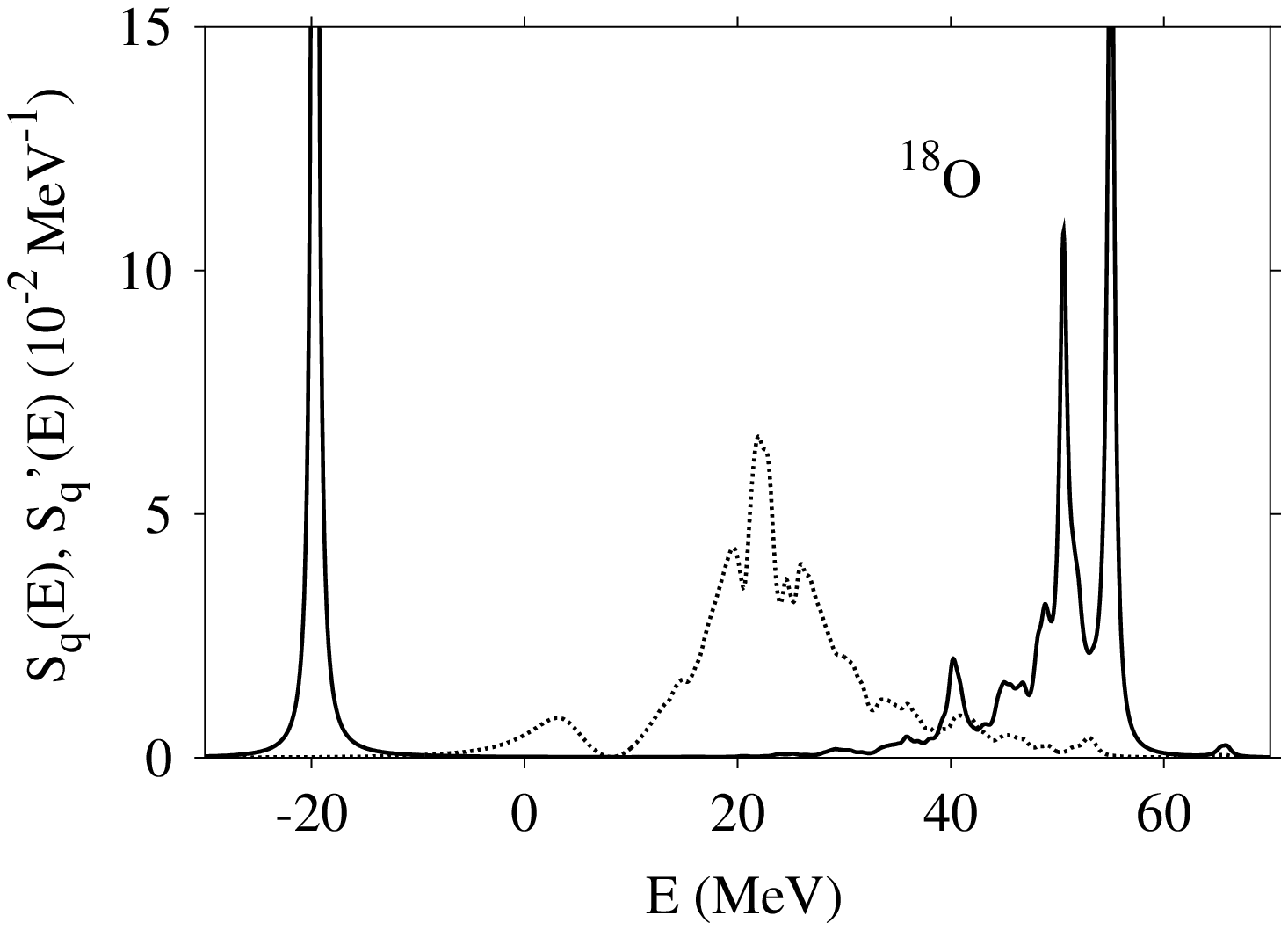,width=85mm,angle=0}
\caption{\label{f1}
Strength functions $S_q(E)$ (solid line) and $S_q'(E)$ (dashed line) 
of PDM phonon distribution in $^{16}$O and $^{18}$O.}
\end{figure}

The results of my calculations \cite{code} are shown in Fig.~\ref{f1}.
A difference in $S_q(E)$ for $^{16}$O and $^{18}$O in Fig.~\ref{f1} is
dramatic but not surprising. It is due to the fact that a phonon couples
to all $1p1h$ configurations with an equal strength of $F_1=1.025$~MeV 
(a PDM parameter) in $^{16}$O
while a coupling strength for $[1p_{1/2} 2d_{5/2}]_{\nu}$ 
configuration at $E_{[1p_{1/2} 2d_{5/2}]_{\nu}}=8.2$~MeV 
in $^{18}$O has been enhanced to 
$F_1'= 40 F_1 = 41$~MeV 
following the details of calculations in Ref.~\cite{n1}.
As a result,
we find the GDR in $^{18}$O between 40 and 60 MeV and about 40\% of its
strength is pushed to $-20$~MeV.
The energy of the ground state is 0~MeV.

Comparing dashed lines in Fig.~\ref{f1}, one may be surprised that $S_q'(E)$ 
does not feel this enormous matrix element of 41~MeV in $^{18}$O
confirming the results in \cite{n1}.
But a very important detail is that
$S_q'(E)$ in Fig.~\ref{f1} has been calculated with $E_{GDR} = 22.5$~MeV 
from Ref.~\cite{n1}.
It has been
done in an attempt to reproduce the GDR strength functions published in Fig.~3
of \cite{n1} which are found in agreement with the data available in 
$^{16}$O and $^{18}$O.
Taking into account that the employed $1p1h$ spectrum might be not exactly 
the same as in Ref.~\cite{n1}, it is possible to conclude that 
dashed curves in Fig.~\ref{f1} reproduce the results in Fig.~3a,~3b 
of Ref.~\cite{n1} on a rather good qualitative level.

\begin{figure}
\epsfig{figure=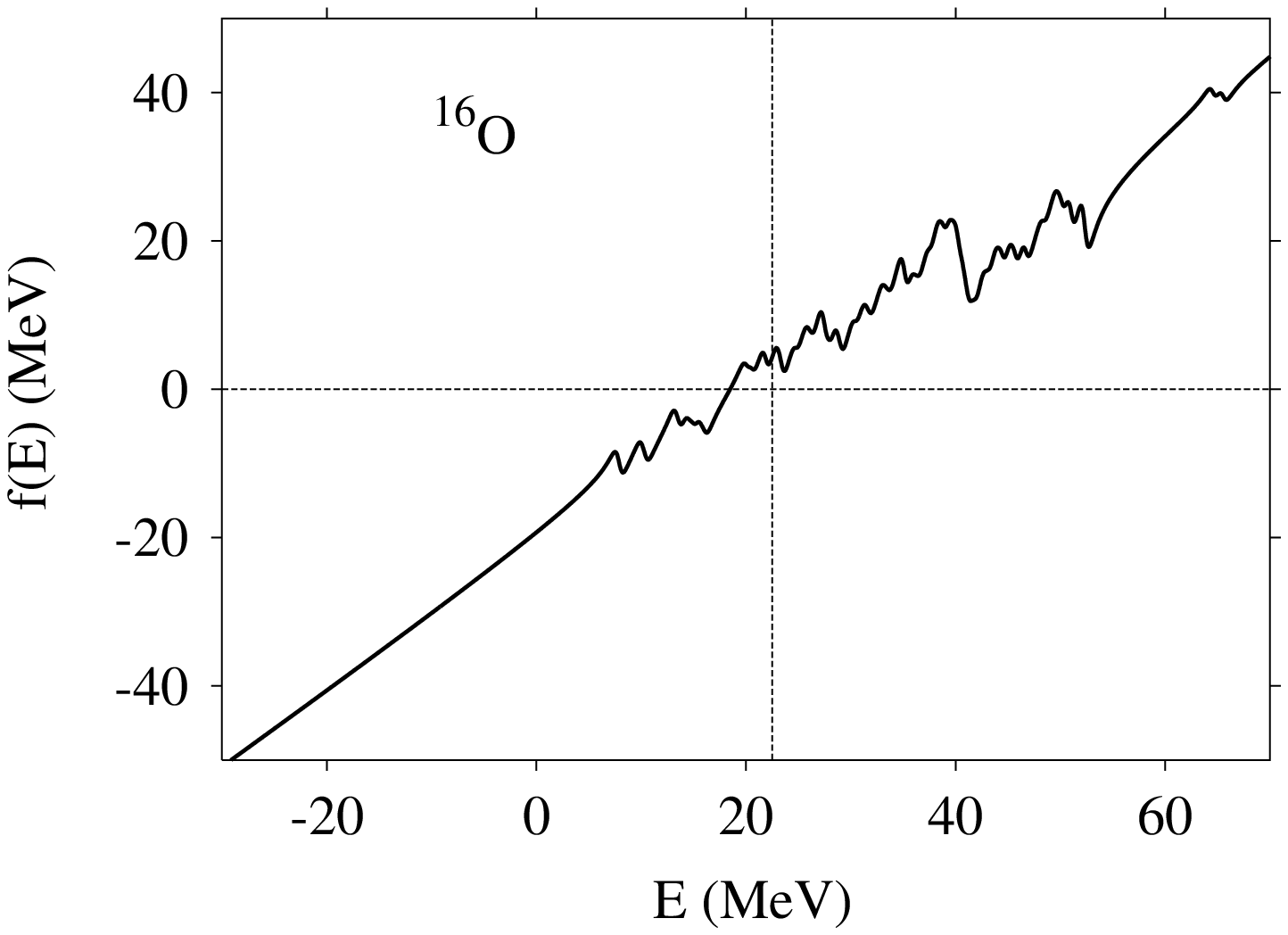,width=85mm,angle=0}
\epsfig{figure=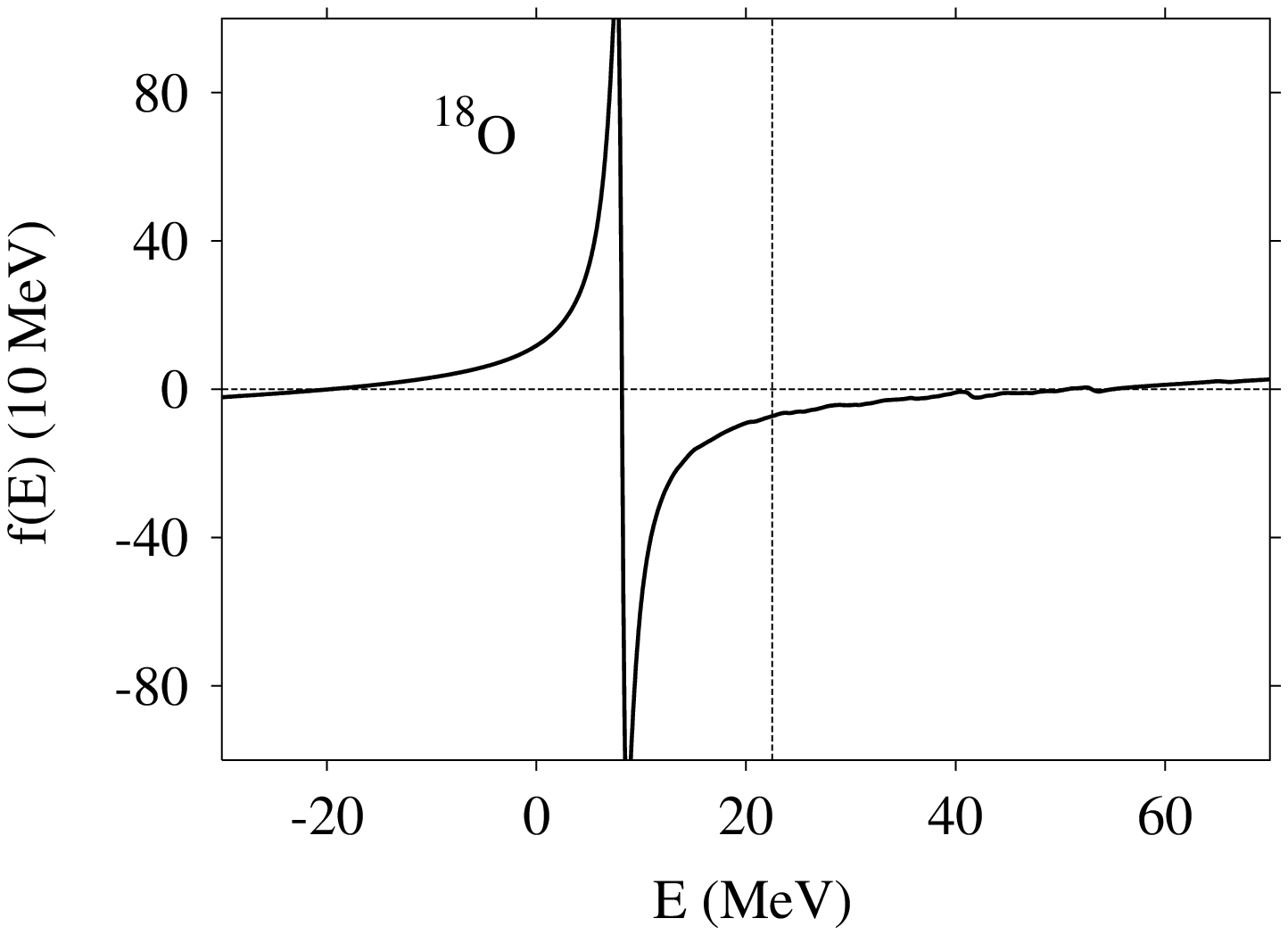,width=85mm,angle=0}
\caption{\label{f2}
Functions $f(E)$ in $^{16}$O and $^{18}$O.
Vertical line is plotted at 22.5~MeV.}
\end{figure}

The only problem is that it is not possible to obtain
$E_{GDR} = 22.5$~MeV in $^{18}$O, reported in Ref.~\cite{n1}, 
with parameters from this article as a solution of Eq.~(\ref{e3}).
To demonstrate this, let us consider a behavior of
the function $f(E)$ of Eq.~(\ref{e3}) in $^{16}$O and $^{18}$O.
In $^{16}$O, it has a tendency of a continues increase
with fluctuations reflecting $1p1h$ poles smeared by the parameter
$\varepsilon$ and crosses $y=0$ line in my calculation
at $E_{GDR} = 18.5$~MeV (see, Fig.~\ref{f2}, left).
In $^{18}$O (Fig.~\ref{f2}, right), a fluctuation around 
$E_{[1p_{1/2} 2d_{5/2}]_{\nu}}$ increases
enormously because of 41~MeV coupling matrix element corresponding to this
pole yielding a spurious solution of Eq.~(\ref{e3}) at this energy. 
The physical PDM solutions
of Eq.~(\ref{e3}) in $^{18}$O have $E_{GDR} = -19.7$, 50.6, and 55.0~MeV
which are very different from $E_{GDR} = 22.5$~MeV, reported in 
Ref.~\cite{n1}. $S_q'(E)$ calculated with any them  
is dramatically different from the one in Fig.~\ref{f1} (right).

The ``$^{18}$O effect" can be obtained even without any calculations.
For that, one may neglect a phonon coupling to all $1p1h$ configurations
(with a ``weak" matrix element $F_1$) except for $[1p_{1/2} 2d_{5/2}]_{\nu}$
configuration. Then, in the limit $\varepsilon \to 0$, Eq.~(\ref{e3})
transforms into quadratic equation: 
\begin{equation} 
E-\omega_{q}-\frac{(F_1')^2 \cdot n}{E-E_{[1p_{1/2} 2d_{5/2}]_{\nu}}}=0~.
\label{e4}
\end{equation} 
where a factor $n$ accounts partial occupation of $\nu2d_{5/2}$ level 
in $^{18}$O. 
Eq.~(\ref{e4}) yields the PDM eigen states at $-19.0$ and 49.4~MeV with 
a phonon strength distribution among them as 40\% and 60\%, respectively.

It becomes clear that agreement with experiment for $^{18}$O (and accordingly
for other neutron-rich nuclei) reported in Ref.~\cite{n1}
has been obtained by making use of the approximate PDM strength function 
$S_q'(E)$ and the GDR energy which is not a solution of Eq.~(\ref{e3})
as announced.
Correct PDM strength function with parameters from Ref.~\cite{n1} 
for $^{18}$O is presented by solid curve in right part of Fig.~\ref{f1}.
 
\vspace*{3mm}

{\bf 1 and 10.} I have examined the PDM from the point of view of
general physical grounds. My arguments and conclusions are 
presented in \cite{m} and above.
I think a reader may independently conclude whether general rules are not for 
this model (as the claims of Dang {\it et al.} in \cite{nn} may be
understood) and whether one learns any physics from the PDM fits in 
Refs.~\cite{n9,n1,o1,n2,n3,n4,n5,n6,n7,n8,n10,n11,n12,n13,n14,n15,n16}
although agreement with experiment is always reported by the authors.


\begin{thebibliography}{99}
\bibitem{code}
Calculations have been performed with a Fortran code pdm.f of 42 lines.
It is available from the author. Calculation time of both $S_q(E)$ and 
$S_q'(E)$ is about 1 sec. on 166~MHz PC.

\vspace*{3mm}
\end{thebibliography}
\end{document}